\documentclass[english,superscriptaddress,twocolumn,PRE]{revtex4-1}
\usepackage{babel}
\usepackage{docs}
\usepackage{bm}
\usepackage[colorlinks=true,linkcolor=blue]{hyperref}
\usepackage{color}
\usepackage[usenames,dvipsnames]{xcolor}
\usepackage{graphicx}
\usepackage{dcolumn}
\usepackage{natbib}
\usepackage{subfigure}
\usepackage{relsize}
\usepackage{amssymb,amsmath}
\usepackage{epstopdf}
\usepackage{wrapfig}
\usepackage{float}
\usepackage{yfonts}
\usepackage{todonotes}
\usepackage{setspace}
\usepackage{latexsym}
\usepackage{graphicx,color}
\usepackage[utf8]{inputenc}
\usepackage[T1]{fontenc}    



\definecolor{DarkRed}{rgb}{.9,0,0}

\usepackage{graphicx,color}

\begin{document}
\title{Interfacial fluid instabilities and Kapitsa pendula}

\author{Madison S. Krieger}
\email{mkrieger@fas.harvard.edu}
\affiliation{School of Engineering, Brown University, Providence, RI 02912, USA}
\affiliation{Program for Evolutionary Dynamics, Harvard University, Cambridge, MA 02138 USA}
\thanks{We acknowledge several motivating and helpful discussions with Mark Levi and Gerhard H. Wolf.}

\date{\today}

\begin{abstract}
The onset and development of instabilities is one of the central problems in fluid mechanics. Here we develop a connection between instabilities of free fluid interfaces and inverted pendula. When acted upon solely by the gravitational force, the inverted pendulum is unstable. This position can be stabilized by the Kapitsa phenomenon, in which high-frequency low-amplitude vertical vibrations of the base creates a fictitious force which opposes the gravitational force. By transforming the dynamical equations governing a fluid interface into an appropriate pendulum-type equation, we demonstrate how stability can be induced in fluid systems by properly tuned vibrations. We construct a  ``dictionary"-type relationship between various pendula and the classical Rayleigh-Taylor, Kelvin-Helmholtz, Rayleigh-Plateau and the self-gravitational instabilities. This makes several results in control theory and dynamical systems directly applicable to the study of tunable fluid instabilities, where the critical wavelength depends on the external forces or the instability is suppressed entirely. We suggest some applications and instances of the effect ranging in scale from microns to the radius of a galaxy. 
\end{abstract}

\maketitle

\section{Introduction}
Interfacial instabilities are the seeds of many pattern-forming mechanisms in Nature\cite{crosshoh}. The ability to delay or entirely suppress these instabilities therefore represents a significant tool. In fluid mechanics, a number of important instabilities occur at one or more discrete interfaces between immiscible fluids with different material and flow properties \cite{Chandrasekhar,Drazin}. Much recent attention has been paid to potential methods of stabilizing these interfaces. Several mechanisms are available; for example, stabilization of the canonical Rayleigh-Taylor instability \cite{kull1991} has been proposed through the use of gyroscopic forces \cite{bsh2015}, magnetically-charged colloids \cite{prr2013}, or heat and mass transfer across the interface \cite{xe2006}. Similarly, capillary instabilities have been shown to be tunable in the presence of internal flows \cite{rs1989}, external acoustic waves \cite{mltm1997,mltm2001,bwbd2010}, and vibrations \cite{lc1991,preeeee,Shevtsova_Gaponenko_Yasnou_Mialdun_Nepomnyashchy_2016}. Tuning the properties of these instabilities is desirable, for example, in the design of fusion reactors. In such systems it can be desirable to find criteria for global stability which do not depend on feedback mechanisms, but rather depend on inducing a tunable external force which causes the unperturbed/trivial solution to be stable for long times.

In classical mechanics, an external agent can always induce such a force according to d'Alembert's principle, by accelerating the previously-inertial frame in which the desired solution was unstable \cite{ArnoldMMCM}. A canonical example is the Kapitsa phenomenon, in which a pendulum in the inverted position can be stabilized by low-amplitude, high-frequency vertical vibrations of its base. This effect was predicted based on variants of the Mathieu equation by \cite{Stephenson1908,Stephenson1909} and explained in depth by \cite{kapitsa1965a,kapitsa1965b,Chelomei1983}. The motion of the pendulum in the presence of the vibration is described by Mathieu's equation. The stability of the trivial solution is determined by the material parameters and the parameters of the oscillator, and several stable regions exist in this parameter space. Recent work has sought to explain and generalize this effect, using tools ranging from symplectic topology \cite{levi1988} to differential geometry and classical mechanics \cite{levibroer1995,levi2005}. Using external vibrations to transform the equation of a simple harmonic oscillator into a Mathieu-type equation has several extensions. The use of the effect to circumvent Earnshaw's theorem, according to which any stationary collection of electric charges is inherently unstable, led to the development of the ion trap \cite{paul1990} which earned its inventor the Nobel Prize. It is now known that all extrema of a potential become minima when the potential undergoes similar oscillations \cite{levi2005}, leading to vast applications in conservative systems. These oscillations have been invoked as a stabilizing mechanism as far afield as economics \cite{hw2003}. They have also been known to stabilize a denser fluid atop a lighter fluid for many decades, presenting one method of dynamically stabilizing the Rayleigh-Taylor instability \cite{bu1954,troyongruber1971,wolf1969,wolf1970,kosz2005}. A similar effect can levitate a rigid body by placing it in a small-amplitude high-frequency oscillating airflow; this is likely an important mechanism in insect flight \cite{wflcz2010}. 

In this work we expand previous work on stabilizing discrete fluid-fluid interfaces with the Kapitsa effect. By reducing the local dynamics of the interface to a one-dimensional dynamical equation for the perturbation amplitude via standard techniques in the study of instabilities, we reveal that many free-boundary instabilities --- including the Rayleigh-Taylor, Kelvin-Helmholtz, Rayleigh-Plateau and self-gravitational instabilities --- can be expressed in terms of a simple harmonic oscillator with a spring constant that is either positive or negative, depending on the relationship of the perturbation wavenumber to the critical wavenumber of the instability. These equations are identical to the linearization of an inverted or standard pendulum, respectively; therefore it is expected that applied vibrations will also transform the dynamics of fluid-fluid interfaces into Mathieu-type equations. Using this correspondence, we give approximate bounds to the largest and most accessible Mathieu-stable region for the interface and suggest ways to induce the stabilizing vibrations.

The layout of the paper is as follows: In \S\ref{kapitpensec} we reprise the basic theory of the Kapitsa pendulum and its connection to the Rayleigh-Taylor instability, referring to what is known in the literature. We derive an approximation for the stability bounds when the acceleration due to external oscillations, $c$, is much larger than the gravitational acceleration. We invoke analyses of \cite{ArnoldMMCM,landau_lifshitz_mech} to derive the conditions of stability. We expand a perturbation of an infinite two-dimensional interface between two fluids of different densities in normal modes, arriving at a one-dimensional dynamical equation for the interface height which is identical to a simple harmonic oscillator. We expand upon this basic planar geometry in \S\ref{extensionssec}, considering multiple superposed Rayleigh-Taylor interfaces. We also include two extensions of an essentially non-Hamiltonian character. These are the Kelvin-Helmholtz instability, which can be represented as a pendulum with complex damping, and the case of a viscous two-layer system confined in a narrow channel. By arguments of scaling we discover that the Kapitsa effect can no longer be observed in the lubrication limit and discuss the relevant dynamics. Due to their similar cylindrical geometry, we analyse the Rayleigh-Plateau and the self-gravitational instability in \S\ref{cylindersss}, where we discuss possible applications to the study of capillary and astrophysical flows. We conclude in \S\ref{conclusions}.

\section{Review of Kapitsa's pendulum and the Rayleigh-Taylor instability}
\label{kapitpensec}
\subsection{The Kapitsa effect}
We begin with a review of the vertically-vibrated inverted pendulum \cite{ArnoldMMCM,kapitsa1965a,kapitsa1965b,landau_lifshitz_mech}. The pendula are systems of point masses connected to rigid, massless rods of length $l$. When acted on solely by the gravitational force, Newton's second law and the small-angle approximation give the well-known system 
\begin{equation}
\ddot{\theta}=-\frac{g}{l} \sin\theta \approx -\frac{g}{l} \theta, \label{regpend}
\end{equation}
where $\theta$ is measured with respect to the vertical axis. Linearizing $\sin\theta$ about the inverted position is identical to the transformation $g \rightarrow -g$. In the inverted system, the trivial solution $\theta=0$ is unstable to small perturbations.  Under the action of other external forces, however, this equilibrium can be stabilized. Kapitsa \cite{kapitsa1965a,kapitsa1965b} suggested the use of a continuous vertical repositioning of the point mass; if the position of the point mass to be given by $(-l \sin \theta, y + l \cos \theta)$, then the resulting dynamical equation is Hill's equation
\begin{equation}
l \ddot{\theta} - \ddot{y} \sin \theta = g \sin \theta \label{movingupsidependulum}.
\end{equation}
If $y(t) = a \cos(\omega t)$, we can add the d'Alembert acceleration due to the motion to the usual gravitational acceleration $g$, making it a time-dependent function
\begin{equation}
g(t) = g - a \omega^2 \cos (\omega t), \label{gnonaut}
\end{equation}
so that the system is no longer autonomous; the value of the corresponding Hamiltonian is now time-dependent through $g(t)$.  As a consequence (\ref{movingupsidependulum}) becomes Mathieu's equation, 
\begin{equation}
\ddot{\theta} = \left(\frac{g}{l} - \frac{a \omega^2}{l } \cos(\omega t)\right) \theta, \label{matty}
\end{equation}
after the small-amplitude approximation $\sin \theta \approx \theta$ has been invoked. The trivial solution of Mathieu's equation is strongly stable for certain values of $a$ and $\omega$ in the range $a \ll 1$, $\omega \gg 1$. It is noteworthy that although the amplitude of the oscillations is very small, the resulting acceleration $c = a \omega^2$ can be much greater than $g$, so that these must be considered violent vibrations of the pendulum and thus are distinct from the phenomenon of parametric resonance. 

An approximation to the value of $a \omega^2$ for which the inverted position of the pendulum is stabilized can be found using an averaging argument of Landau and Lifshitz \cite{landau_lifshitz_mech}. The angle of the pendulum can be written as $\theta(t)=\Theta(t)+\xi(t)$, where $\xi(t)=a \cos (\omega t)$ represents the displacement due to the forced oscillations, and therefore has small magnitude and mean zero over the period of the forcing, whilst $\Theta(t)$ changes by only small increments over one period of the forcing. Averaging Mathieu's equation, Eq. (\ref{matty}), reveals that the dynamics for a pendulum with angle $\Theta(t)$ are given by an effective potential
\begin{equation}
U_{eff} =  m g l \left[-\cos ^2 \Theta + \left(\frac{a^2 \omega^2}{4 g l}\right) \sin ^2 \Theta \right]\label{arnoldapprox}.
\end{equation}
By examining the minima of this effective potential, one sees that $\Theta=\pi$ is always stable (remembering that the approximation $a \ll 1$ was invoked), and $\Theta=0$ is stable so long as
\begin{equation}
a^2 \omega^2 > 2 gl \label{arnoldcriterion}.
\end{equation}

\begin{figure}
\includegraphics[width=\columnwidth]{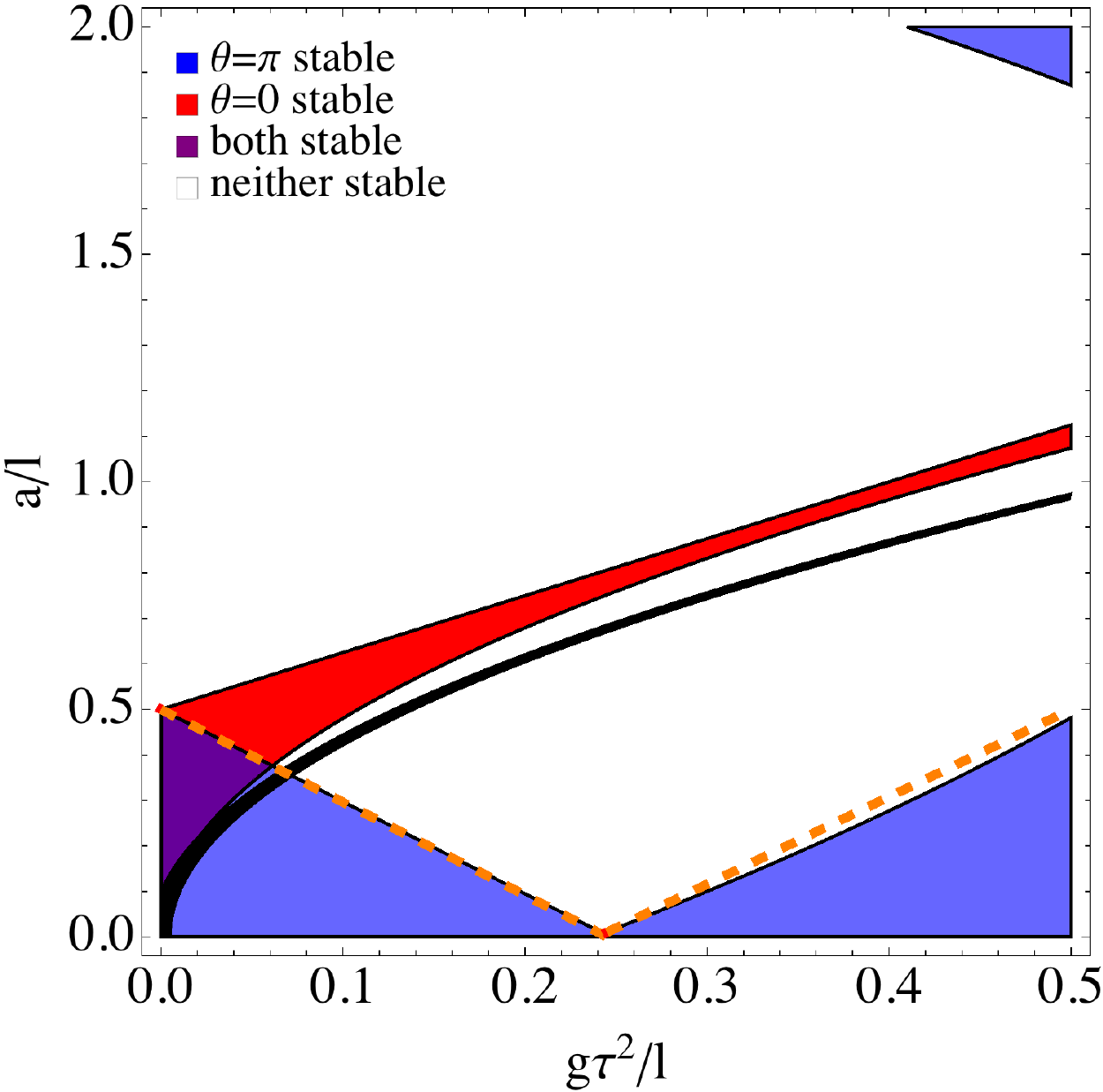}
\caption{A relevant region of the stability diagram for the Mathieu equation, Eq. (\ref{matty}). Here our convention is that $\theta=\pi$ corresponds to the hanging position of a standard pendulum and $\theta=0$ corresponds to the inverted position. The thick black curve is given by Eq. (\ref{arnoldcriterion}), and the orange dotted line is given by Eq. (\ref{resonance}). The rest of the diagram is filled in by adapting the work of \cite{bhnsv2004} to our parameter regime. Although in general stability of this system is very complicated, in this work we focus on the regime $a \ll l$, $g \tau^2 \ll a$ and are therefore mainly interested in the transition from stability of the hanging pendulum (blue) to bistability (purple). } 
\label{stabMatfig}
\end{figure}

However, the effect of external vibration on stability extends beyond the stabilization of the inverted state. For particular parameter values, it leads to an exchange of stability, in which the hanging state of the parameter also becomes unstable. In certain regimes it also leads to destabilization of both states; therefore it is possible for there to be zero, one, or two stable equilibria depending on parameters. The simplest way to observe the destabilization of the hanging state is to treat the system as Hamiltonian with the governing function $H = p^2 /2 - g(t) \cos \theta$. By truncating the Taylor series at $\cos \theta \approx (1-\phi^2/2)$ for long periods of the driving, we recover the Hamiltonian for a driven harmonic oscillator. It is easily checked \cite{landau_lifshitz_mech} that the first parametric resonance in this limit is given by 
\begin{equation}
\frac{2 a \pi^2}{l \tau^2} \leq \left| \frac{\pi^2}{\tau^2}-\frac{4g}{l}\right|. \label{resonance}
\end{equation}
Subsequent resonances occur for more extreme material parameters, leading to a periodic stability diagram with slightly changing shapes in which exchanges of stability take place (Floquet tongues). Finding the entire stability portrait requires more care; details can be found in \cite{ms2004,ms2009,bhnsv2004}. To clarify the basic details in the parameter regime we are interested in, we have provided a stability diagram in Fig. \ref{stabMatfig}. Because we are interested in a region where both $a$ and $\tau$ are very small, we are guaranteed that one solution will be stable, and are therefore primarily interested in the transition from stability of only one solution (the hanging pendulum) to bistability. As shown in the figure, Eq. (\ref{arnoldcriterion}) and Eq. (\ref{resonance}) are sufficiently accurate to understand the first transitions, from stability of the hanging solution via a pitchfork bifurcation \cite{bhvv1998k} to bistability and then from bistability to stability only of the inverted solution, as a function of the external oscillator amplitude and period. 

\subsection{The inviscid Rayleigh-Taylor instability}
\label{RTinst}

\begin{figure}
\centering
\includegraphics[width=\columnwidth]{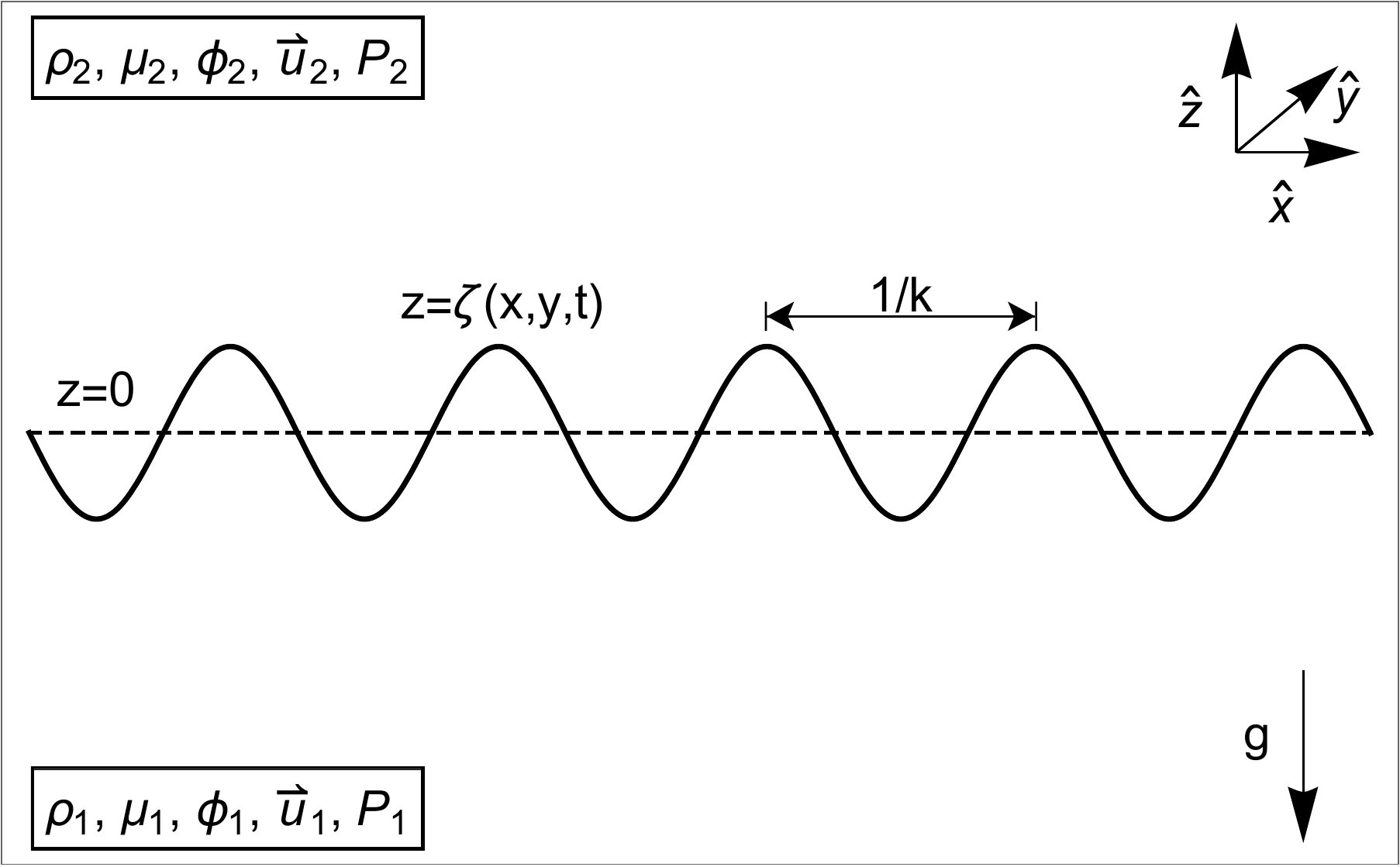}
\caption{The Rayleigh-Taylor instability. Two fluids of differing material properties are superposed and separated by an interface $z = \zeta(x,y,t)$. An expansion in normal modes $k \equiv \sqrt{k_x ^2 + k_y ^2}$ reduces the dimension of the dynamics to one.} 
\label{figgy1}
\end{figure}

We now turn our attention the case of two unbounded, incompressible, immiscible fluids sharing a common interface which is a perturbation of the $z=0$ plane. We label the fluid in the region $z<0$ as fluid $1$ with velocity $\boldsymbol{u}_1$ and density $\rho_1$. We label the fluid in the region $z>0$ as fluid $2$ with velocity $\boldsymbol{u}_{2}$ and density $\rho_{2}$. The shape of the perturbation is given by $\zeta(x,y,t)$. The setup is shown in figure \ref{figgy1}. In this initial calculation we neglect the effects of viscosity and surface tension. We consider a quiescent initial condition, so that the fluid is irrotational for $t>0$. We therefore define a potential, $\boldsymbol{u}_i = \nabla \phi _i$. We assume both fluids are incompressible and that far away from the interface the flow vanishes. On the interface itself we assume $p_2=p_1$; furthermore, since we will be interested primarily in the stability of the trivial solution $\zeta(x,y,t)=0$, we will assume the speeds at the interface are small, so that we can linearize Bernoulli's principle relating the unsteady heads at the interface. Our system is then given by the equations
\begin{eqnarray}
\nabla ^2 \phi _i = 0, \ \ \ \label{laplace} \\
\nabla \phi _i(z \rightarrow \pm \infty) \rightarrow 0,  \ \ \ \label{nofarflow} \\
\partial_z \phi _i |_{z=0} = \partial _t \zeta(x,y,t), \ \ \ \label{intBC} \\
\rho_1(\partial _t \phi_1 |_{z=0} + g \zeta(x,y,t)) = \rho_2(\partial _t \phi_2 |_{z=0} + g \zeta(x,y,t)). \ \ \ \label{bern}
\end{eqnarray}

Because the system is linear, we assume an ansatz of normal modes:
\begin{eqnarray}
\lbrace \phi_1(x,y,z,t), \phi_2(x,y,z,t), \zeta(x,y,t)\rbrace  \nonumber \\
= \lbrace \hat{\phi}_1(z,t), \hat{\phi}_2(z,t), \hat{\zeta(t)} \rbrace e^{i(k_x x + k_y y)}, \label{normmodes}
\end{eqnarray}
where we will often use the variable $k \equiv \sqrt{k_x ^2 + k_y ^2}$. Under this assumption, (\ref{laplace}) becomes $(\partial_z ^2 - k^2)\hat{\phi}_i = 0$, which have the solution $\hat{\phi}_i = C_i(t) e^{\pm k z}$, where the $\pm$ is positive for fluid 1 and negative for fluid 2 by the boundary conditions (\ref{nofarflow}). Substitution of these exponentials into (\ref{intBC}) gives
\begin{equation}
\partial _t \hat{\zeta}(t) = \pm k C_i(t)  e^{\pm k z} \longrightarrow \partial_t \hat{\phi}_i |_{z=0} = \pm \frac{1}{k} \partial _{tt} \hat{\zeta}(t).
\end{equation} 
A final substitution into the Bernoulli relation reveals that the interface height behaves like a simple harmonic oscillator:
\begin{equation}
\partial _{tt} \hat{\zeta} = k A g \hat{\zeta}, \label{RTinv}
\end{equation}
where $A \equiv \frac{\rho_2 - \rho_1}{\rho_2 + \rho_1}$ is the Atwood number. The behavior of this system under the action of external vibrations has been studied in the literature \cite{bu1954,wesson1970}. 

We now assume that $A>0, \ \rho_2>\rho_1$, and that the entire system is made to oscillate in the $\hat{z}$ direction. We assume that the position of every plane of constant $z=z_c$ actually moves according to the form $z_c (t) = a \cos (\pi t/\tau)$.  These types of vertical vibrations of two-fluid systems commonly lead to a host of pattern-forming instabilities known as Faraday waves. Because we are only interested in the dynamics of the interface height, these phenomena are outside the scope of the current work; for an overview of the basic linear theory, see \cite{kt1994}; for recent advances, see \cite{hf2013}. The patterns involved are a subject entire unto themselves, and can be used, for example, to assemble complicated patterns \cite{chenetal2014}. The time-dependent gravitational acceleration in our pendulum-like equation is now given by (\ref{gnonaut}). When $g<c$, (\ref{RTinv}) can then be approximated by the effective potential of (\ref{arnoldapprox}), and stability of the trivial solution $\hat{\zeta}=0$ is defined by a condition analogous with (\ref{arnoldcriterion}):

\begin{equation}
k a^2 \omega^2 > 2 g A. \label{RTcritcrit}
\end{equation}
A similar criterion has been derived in the fluids literature by \cite{troyongruber1971,wolf1969}. Because all wavenumbers $k$ are admissible there is an upper bound $k_{max}$ above which the system passes from the first stable Mathieu band to the second unstable band. This upper bound to the first stability region is given physically by requiring that the frequency of the oscillator is greater than the instability growth rate $\Omega^2$ due to its induced acceleration. This may be written $\Omega^2 < k c A$ and after substitution one finds that this second instability does not occur so long as 
\begin{equation}
1 > A a k. \label{upperbound}
\end{equation}
For a more detailed derivation of this balance, see \cite{wolf1969}.

\subsection{Previous work on extensions to this system}

For material parameters representing common systems such as water suspended above air, or a general liquid suspended above a gas, Eq. (\ref{RTcritcrit}) and Eq. (\ref{upperbound}) do not cover the whole band of disturbance wavenumbers. In practical application, therefore, one needs a mechanism to suppress the very small-wavelength disturbances, such as surface tension or viscosity, as very large-wavelength disturbances are often prohibited by vessel diameter. It has been suggested \cite{troyongruber1971} that for material parameters matching an air-water interface stability for all $k$ is only guaranteed by including both of these effects together; including only one or the other leaves some range of $k$ unstable. A routine calculation demonstrates that the inclusion of surface tension decreases the time average of $g(t)$, widening the range of stable $k$ in both directions. Following the same steps as the previous section but including the pressure difference due to surface tension leads to the harmonic oscillator
\begin{equation}
\hat{\zeta}_{tt} = \left(k A g - \frac{\gamma k^3}{\rho_1 + \rho_2}\right) \hat{\zeta}, \label{surteneqn111}
\end{equation}
For fluids with surface tension an order of magnitude larger than water, this can be sufficient to stabilize the entire range of $k$. Such fluids were likely not considered by the authors of \cite{troyongruber1971}. 

Considering a viscous fluid leads to the dynamical equation for the interface matching a damped harmonic oscillator, where the value of the damping depends on the kinematic viscosities $\nu_2, \ \nu_1$ of the two fluids as well as some numerical prefactors. For the right damping parameters this can stabilize large-$k$ disturbances; however, too much damping causes the restoring force due to the Kapitsa effect to be damped out as well, causing an instability. This is often referred to as a ``secondary dissipative instability''; for the general theory, see \cite{km2006}.

We conclude on a brief note concerning the possibility of replacing the external vibrations generating the Kapitsa effect with internal pressure waves. This can be desirable in large-scale applications where it is physically infeasible to translate a fluid-fluid interface or apparatus even with small-amplitude vibrations. We studied such systems in the context of the Rayleigh-Taylor instability, using this framework of comparing linear instabilities to simple harmonic oscillators with negative stiffnesses. Our findings suggest that for gravitationally-driven instabilities, such a method is not feasible. This can be seen from dimensional analysis; if the magnitude of a unimodal pressure wave is $\bar{P}$ at the fluid-fluid interface, then the acceleration of the interface is given by $\frac{\bar{P}}{\rho_1 + \rho_2}$. For an air-water interface, the pressures must be on the order of a few $atm$ at the interface to begin to achieve stabilization. On the other hand, in capillary-driven instabilities where the accelerations are much smaller than $g$, such acoustic waves ``tuned'' to the wavelength of a developing instability have been shown to stabilize the interface \cite{mltm1997,mltm2001,bwbd2010}. We address similar ideas in \S \ref{cylindersss}.

\section{Extensions for planar interfaces}
\label{extensionssec}

\subsection{Coupled interfaces}

\begin{figure}
\includegraphics[width=\columnwidth]{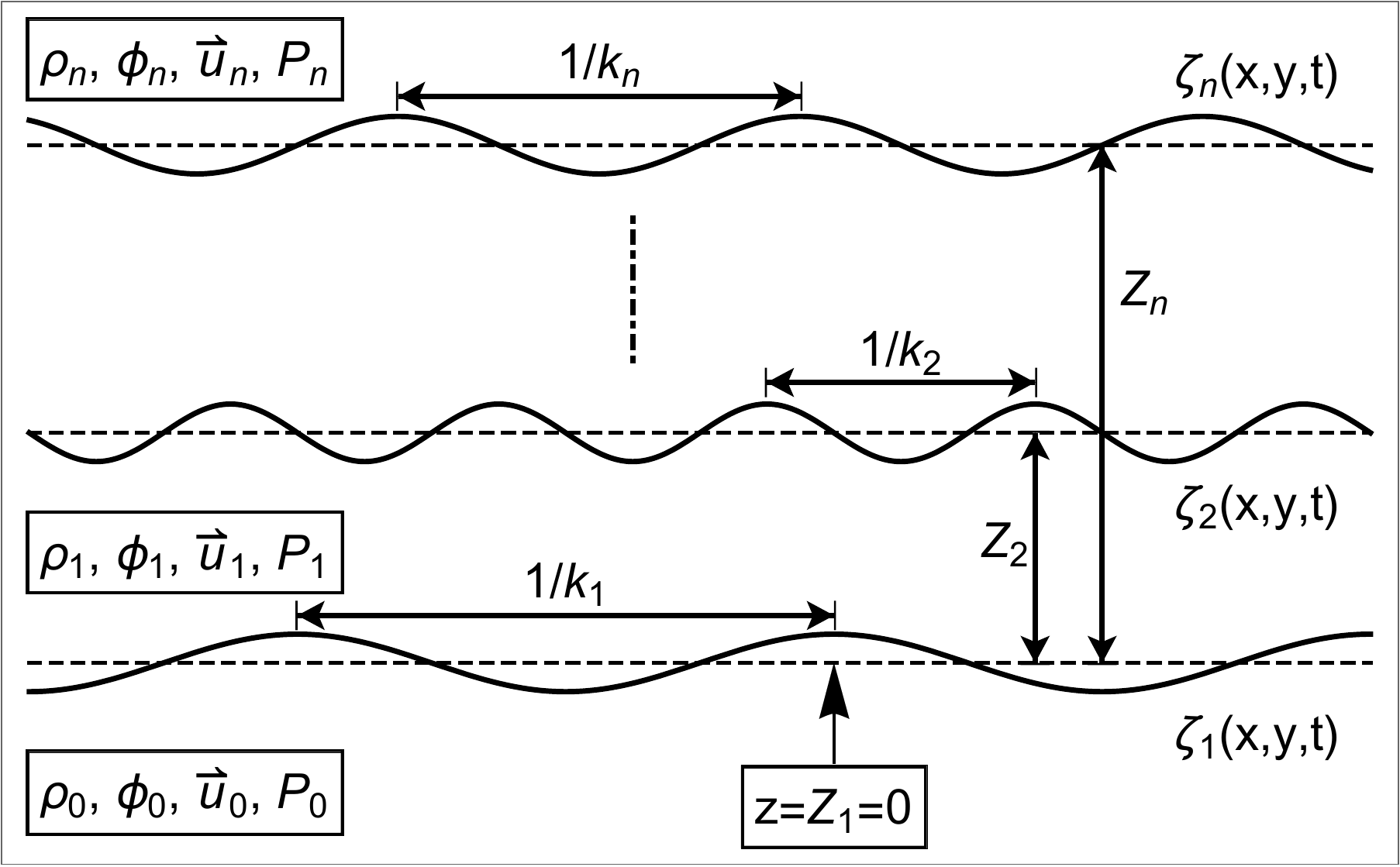}
\caption{Multiple free interfaces between fluid layers of differing density.} 
\label{multpen}
\end{figure}

To see if there is an extension of the interface-pendulum analogy to multiple bodies, we consider a system of $n$ interfaces between $n+1$ fluids of differing density $\rho_i$. We will enumerate the fluid regions from $i = 0$ to $i = n$ and their separating interfaces from $i = 1$ to $i = n$ centered around planes $z = Z_i$, with $Z_1 = z = 0$, as shown in figure \ref{multpen}. The reduced flow potential $\phi_i$ in the bulk regions is described by
\begin{eqnarray}
 (\partial_z^2 - k_i ^2 - k_{i+1}^2)\phi_i = 0, \\
\phi_0 = A_0 e^{k_1 z}, \\
\phi_i = A_i e^{k_{i+1}(z-Z_{i+1})} + B_i e^{-k_i (z-Z_i)}, \\
\phi_n = B_n e^{-k_n (z-Z_n)},
\end{eqnarray}
where the $A_i, B_i$ are linked recursively by continuity at the interfaces, $\partial_z \phi _{i-1} |_{z=Z_i} = \partial_z \phi _{i} |_{z=Z_i}$:
\begin{eqnarray}
A_0 k_1 = -B_1 k_1 + k_2 A_1 e^{-k_2  Z_2}, \ \ \ \ \ \ \ \ \ \ \ \label{recr1} \\
A_i k_{i+1} - B_i k_i e^{-k_i (Z_{i+1} - Z_i)} = A_{i+1} k_{i+2} e^{k_{i+2} (Z_i - Z_{i+1})} - B_{i+1} k_{i+1}, \label{recr2} \\
-B_n k_n = -B_{n-1} k_{n-1} e^{k_{n-1}(Z_{n-1} - Z_n)} + k_n A_n. \ \ \ \ \ \label{recr3}
\end{eqnarray}
These equations cannot be solved if, for example, $k_i = 0 \ \forall i \neq 1$ --- we therefore restrict our attention to the case where all appearing quantities are nonzero. Furthermore, for the interfaces to be discrete, we require $\rho_{i-1} \neq \rho_i$. We can write the equation for the dynamics of the $i$-th interface using the Bernoulli relation
\begin{eqnarray}
\rho_{i-1}(\partial _t \phi_{i-1} |_{z=Z_i} + g' \zeta_i) &=& \rho_i(\partial _t \phi_i |_{z=Z_i} + g' \zeta_i). \label{multbernoulli}
\end{eqnarray}
These considerations lead to the matrix formulations for the inner interfaces $i \neq \lbrace 1,n \rbrace$
\begin{eqnarray}
\partial _t \zeta_i(x,y,t) &=& k_i A_{i-1} - k_{i-1} B_{i-1} e^{k_{i-1} (Z_i - Z_{i-1})} = ... \nonumber \\
&=& K^A _{ij} A_j + K^B _{ij} B_j, \label{mmat1}\\
\phi_{i-1}|_{z = Z_i} &=& A_{i-1} + B_{i-1} e^{-k_{i-1}(Z_i-Z_{i-1})} = ... \nonumber \\
&=& A_{i-1} + F^B _{ij} B_j, \label{mmat2}\\
\phi_i |_{z=Z_i} &=& A_i e^{k_{i+1}(Z_i - Z_{i+1})} + B_i = B_i + F^A _{ij} A_j. \ \ \label{mmat3}
\end{eqnarray}
Similar equations exist for $i=1$ and $i=n$, but for convenience in developing a simple heuristic we restrict our attention to the inner interfaces. Because the quantities appearing in (\ref{mmat1}--\ref{mmat3}) are nonvanishing for generic nonzero parameters, $K_{ij} ^A, \ K_{ij}^B, \ F_{ij} ^A, \ F_{ij} ^ B$ are all generically invertible matrices \cite{ArnoldMMCM}. Furthermore, because (\ref{recr1}--\ref{recr3}) represent a linear recursion system, we may assume $A_j, \ B_j$ are of the form $A_i = R^A _{ij} N_j, \ B_i = R^B _{ij} N_j$, where $N_j$ is a vector comprised of $n$ data closing the system which could include both wavenumbers for certain interfaces and $A$'s, $B$'s, for certain fluid regions. It follows that these solutions are uniquely determined and $R^A_{ij}, R^B _{ij}$ must also be invertible. Because by our assumption all $A$'s and $B$'s are nonzero, $R^A_{ij}, R^B_{ij}$ have no nilpotent part. Because sums and products of invertible matrices are invertible, we can write
\begin{equation}
\partial _t \zeta_i = (K^A _{ik} R^A _{kj} + K^B _{ik} R^B _{kj}) N_j \longrightarrow N_j = S^{-1}_{ij} \partial _t \zeta _i, \label{dztmultpen}
\end{equation}
where $S_{ij}  \equiv  (K^A _{ik} R^A _{kj} + K^B _{ik} R^B _{kj})$. Substitution of (\ref{dztmultpen}) into (\ref{mmat1}--\ref{mmat3}) and subsequently into (\ref{multbernoulli}) puts Bernoulli's law in the form
\begin{eqnarray}
(\rho_{i-1}-\rho_i) g \zeta_i = (\rho_i (R^B_{ik} + F^A_{il} R^A_{lk})S^{-1}_{kj}  \nonumber \\
-\rho_{i-1} (L_{il} R^A_{lk} + F^B_{il} R^B_{lk})S^{-1}_{kj})\partial _{tt} \zeta_j. \label{masterrope}
\end{eqnarray}
The operator in (\ref{masterrope}) appears like a stiffness matrix in the theory of coupled oscillators, and due to our assumptions above it has no nilpotent part and can therefore it can be diagonalized and the problem studied in terms of normal modes in a manner completely analogous to coupled pendula. When multiple pendula are inverted, if $n$ is very large, the model is like a linearly-elastic model of a vertical rope. As a consequence, this problem is sometimes referred to as the ``Indian rope trick''. \cite{Stephenson1908} predicted in his original work on this effect that the stabilization could be extended to multiple coupled pendula; this was demonstrated experimentally by \cite{am1993} and studied in greater detail by \cite{ach1993}. The stability diagram for $n$ coupled inverted pendula (and by extension, the diagram for $n+2$ fluid interfaces whose density increases monotonically with $z$) is given by a series of nested Mathieu stability regions. Because the stability diagram for $n$ inverted pendula is a strict subset of the stability diagram for $n-1$ inverted pendula, the size of the stable region vanishes rather quickly as $n \rightarrow \infty$. One cannot stabilize a large collection of inverted pendula using the Kapitsa effect, and therefore one cannot stabilize a large number of successively dense, discrete fluid regions in regimes where the approximations of our calculation hold.

However, the same group later demonstrated that an \textit{actual} elastic rod which is longer than the self-buckling threshold \textit{can} be stabilized in the straight position by the Kapitsa effect \cite{mcfga2003}. In fact, the rod is stabilized at much lower frequencies than can be predicted from a linear theory. The work of \cite{fc2002} revealed that nonlinear resonances between the driving of the external oscillator and higher-order buckling modes were responsible for this enhanced stabilization. Due to the number of nonlinearities excluded in our model it is entirely possible that such resonances can enhance the stability of our multiple-interface system --- a promising analogue with the geometric nonlinearities present in the elastic rod model may be Bragg resonances between distant interfaces \cite{aly2009}, but consideration of nonlinear effects is outside the scope of the present work.

\subsection{Inviscid Kelvin-Helmholtz instability}

\begin{figure}
\centering
\includegraphics[width=\columnwidth]{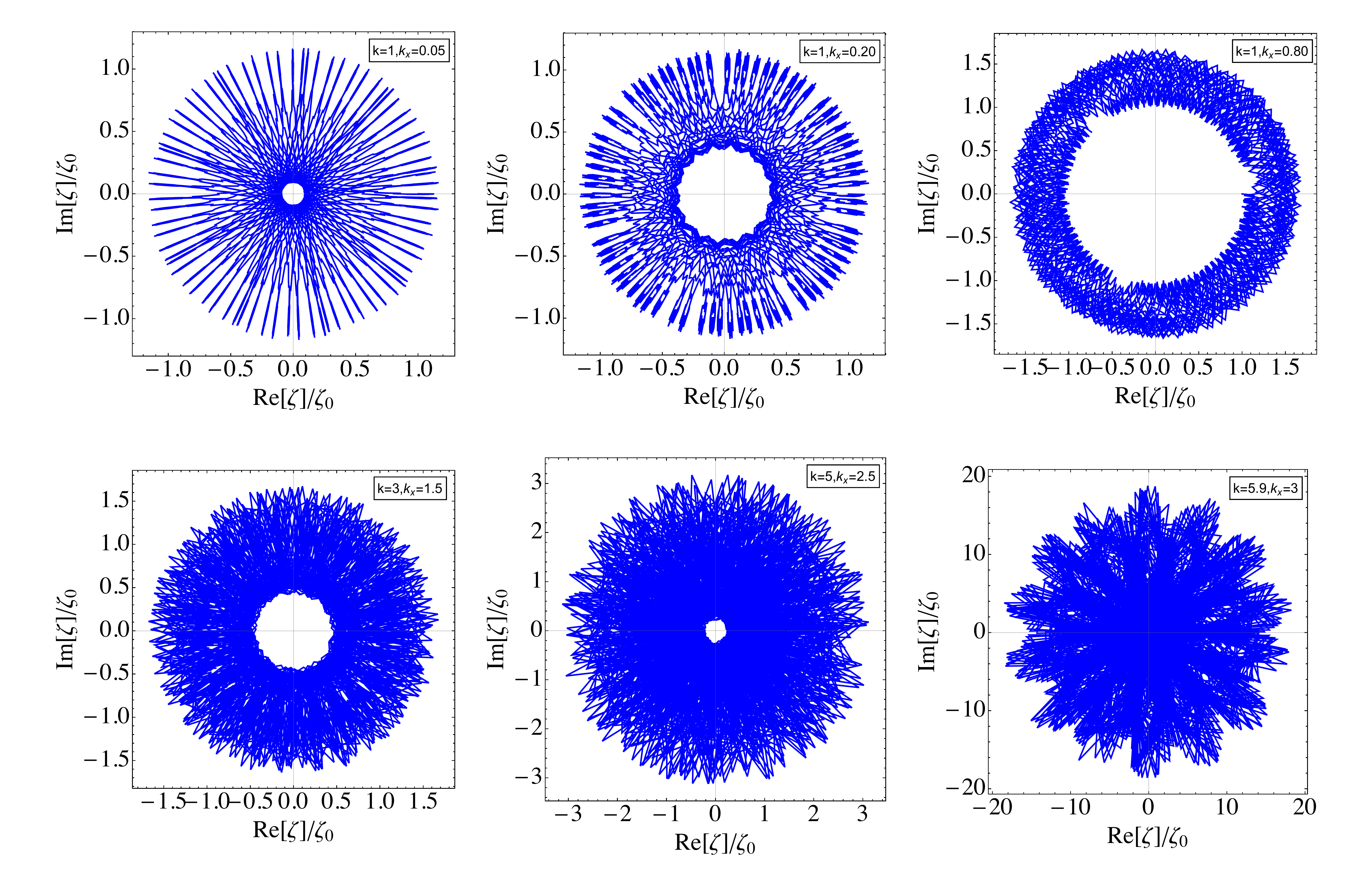}
\caption{Numerical solution of (\ref{KHKHKH1}--\ref{KHKHKH2}) for $A=1$, $U_1 = 0$, $U_2 = 5 \ ms^{-1}$, $\rho_1 \approx 0$, $g(t) = g+(0.08)(70)^2 \cos(t/70)$, and an initial perturbation $\zeta_0 = Z_r(t=0) = 5 \ cm$. The top row represents $k=1 \  m^{-1}$ and increasing $k_x$, plotted over several thousand periods of the oscillator. The bottom row represents $k_x = k/2$ and increasing $k$, plotted over several thousand periods of the oscillator. Instability occurs between $k_{min} \approx 0.638 \ m^{-1}$ and $k_{max} \approx 5.930 \ m^{-1}$, which is roughly half the size of the stability region defined by (\ref{RTcritcrit}-\ref{upperbound}). } 
\label{khplot1}
\end{figure}
We continue our analysis of the problem described in \S\ref{RTinst}, permitting the two layers to have different relative streaming velocities tangential to the interface. For simplicity we will assume both streaming velocities are in the $\boldsymbol{\hat{x}}$ direction and that the fluid is inviscid and irrotational. We assume these velocities $\boldsymbol{u}_i = \left(U_i+u_i,v_i,w_i\right)$ are large enough that $g |\zeta| \ll U_i ^2$; in this limit we write potential as $\phi_i = U_i x + \phi'_i$ and neglect products of $\phi'_i$, following \cite{Drazin}. The instability due to the relative streaming velocities at the interface is a common first instability in transitions to turbulence, and is also responsible for pattern formation in free ocean waves and clouds \cite{jeffreys1925}. We take the streaming velocities to be uniform functions in the $\boldsymbol{\hat{x}}$ direction. The boundary condition on $z = \zeta(x,y,t)$ must be amended to account for the streaming velocity. After the expansion in normal modes (\ref{normmodes}), the system may be described by
\begin{eqnarray}
(\partial_z^2 - k^2)\hat{\phi}'_i = 0, \ \ \ \ \label{kh11}\\
\partial_z \hat{\phi}'_i (z \rightarrow \pm \infty) \rightarrow 0, \ \ \ \ \label{kh12} \\
\partial_z \hat{\phi}'_i |_{z=0} = \hat{\zeta}_t + \mathrm{i} k_x U_i \hat{\zeta}, \ \ \ \ \label{kh13} \\
\rho_1 (\mathrm{i} k_x U_1 \hat{\phi}'_1 + \partial_t \hat{\phi}'_1 + g \hat{\zeta})|_{z=0} = \rho_2 (\mathrm{i} k_x U_2 \hat{\phi}'_2 + \partial_t \hat{\phi}'_2 + g \hat{\zeta})|_{z=0}. \label{kh14}
\end{eqnarray}
A solution to (\ref{kh11}-\ref{kh13}) is given by
\begin{eqnarray}
\hat{\phi}'_j &=& \pm \frac{(\hat{\zeta}_t + \mathrm{i} k_x \hat{\zeta} U_j)}{k} e^{\pm k z}. 
\end{eqnarray}
with the positive sign for fluid 1 and negative for fluid 2. Substitution into (\ref{kh14}) gives the dynamical equation
\begin{equation}
\hat{\zeta}_{tt} + 2 \mathrm{i} k_x \hat{\zeta}_t \frac{\rho_1 U_1 + \rho_2 U_2}{\rho_1 + \rho_2} = \hat{\zeta}\left(k g(t) A + k_x ^2 \frac{\rho_1 U_1 ^2 + \rho_2 U_2 ^2}{\rho_1 + \rho_2}\right). \label{khpendueqn}
\end{equation}
Because of the complexified damping coefficient, the Kelvin-Helmholtz instability represents a type of \textit{self-oscillator}, which unlike a standard linear oscillator can transfer energy between modes; for a review of this topic including a brief note on the instability itself, see \cite{jenkins2013}. Writing $\hat{\zeta}(t) = Z_r(t) + \mathrm{i} Z_i(t)$ and interpreting the dynamics of the actual interface as $Z_r(t)$, (\ref{khpendueqn}) can be rewritten as 
\begin{eqnarray}
\ddot{Z}_r - \frac{2 k_x (\rho_1 U_1 + \rho_2 U_2)}{\rho_1 + \rho_2} \dot{Z}_i  = \left(k g A + k_x ^2 \frac{\rho_1 U_1 ^2 + \rho_2 U_2 ^2}{\rho_1 + \rho_2}\right) Z_r, \nonumber \\
\label{KHKHKH1}\\
\ddot{Z}_i + \frac{2 k_x (\rho_1 U_1 + \rho_2 U_2)}{\rho_1 + \rho_2} \dot{Z}_r  =\left(k g A + k_x ^2 \frac{\rho_1 U_1 ^2 + \rho_2 U_2 ^2}{\rho_1 + \rho_2}\right) Z_i. \nonumber \\  \label{KHKHKH2}
\end{eqnarray}
At first sight, this system cannot be diagonalized over the reals due to the antisymmetric operator coupling the real and imaginary parts of $Z$, and therefore the system looks nothing like a standard oscillator. However, the system is identical to some gyroscopic systems, included coupled and oscillating gyroscopes\cite{rtv1993,bhrw2004}. In fact, motivated by the treatment of these examples in the literature, if we rewrite Eq. (\ref{khpendueqn}) with the shorthand
\begin{equation}
\hat{\zeta}_{tt} -2  \mathrm{i} \alpha \hat{\zeta}_t = \left(k g(t) A+\beta \right) \hat{\zeta},
\end{equation}
and employ the transformation $\zeta = v e^{\mathrm{i} \alpha t}$, we find a single Mathieu equation over the real numbers, 
\begin{equation}
v_{tt}=\left(k g(t) A + \beta - \alpha^2 \right) v.
\end{equation}
Since $\alpha =  k_x (\rho_1 U_1 + \rho_2 U_2) (\rho_1+\rho_2)^{-1}$ is a strictly positive real quantity, stability of $v$ implies stability of $\zeta$ (equivalently, writing out the equations for real and complex parts of $v$ give identical Mathieu equations). This equation appears strongly like Eq. (\ref{surteneqn111}), that for an interface with surface tension. However, the surface tension term is now given by
\begin{equation}
\beta-\alpha^2 = k_x ^2 (U_1 - U_2)^2 \frac{\rho_1 \rho_2}{(\rho_1 + \rho_2)^2},
\end{equation}
which is strictly positive, and therefore the sign is opposite that of standard surface tension (i.e., increases the apparent gravitational acceleration in the pendulum context). Any amount of horizontal streaming therefore has a detrimental effect on stability via the Kapitsa effect. 

To understand the effect of self-oscillation due to streaming, as well as the appearance of a complex-valued interface height, we also solve (\ref{KHKHKH1}--\ref{KHKHKH2}) numerically and find the first stable band of wavenumbers for fixed oscillator parameters and streaming velocities. Some results are shown in Fig.  \ref{khplot1}. For the parameters explored, the size of the stability region in $k$ shrunk monotonically with increasing relative streaming velocity. The size of the stability region in $k$ is approximately half of that predicted by \S\ref{RTinst} for reasonable material parameters.

\subsection{Rayleigh-Taylor instability in superposed thin films}

\begin{figure}
\includegraphics[width=\columnwidth]{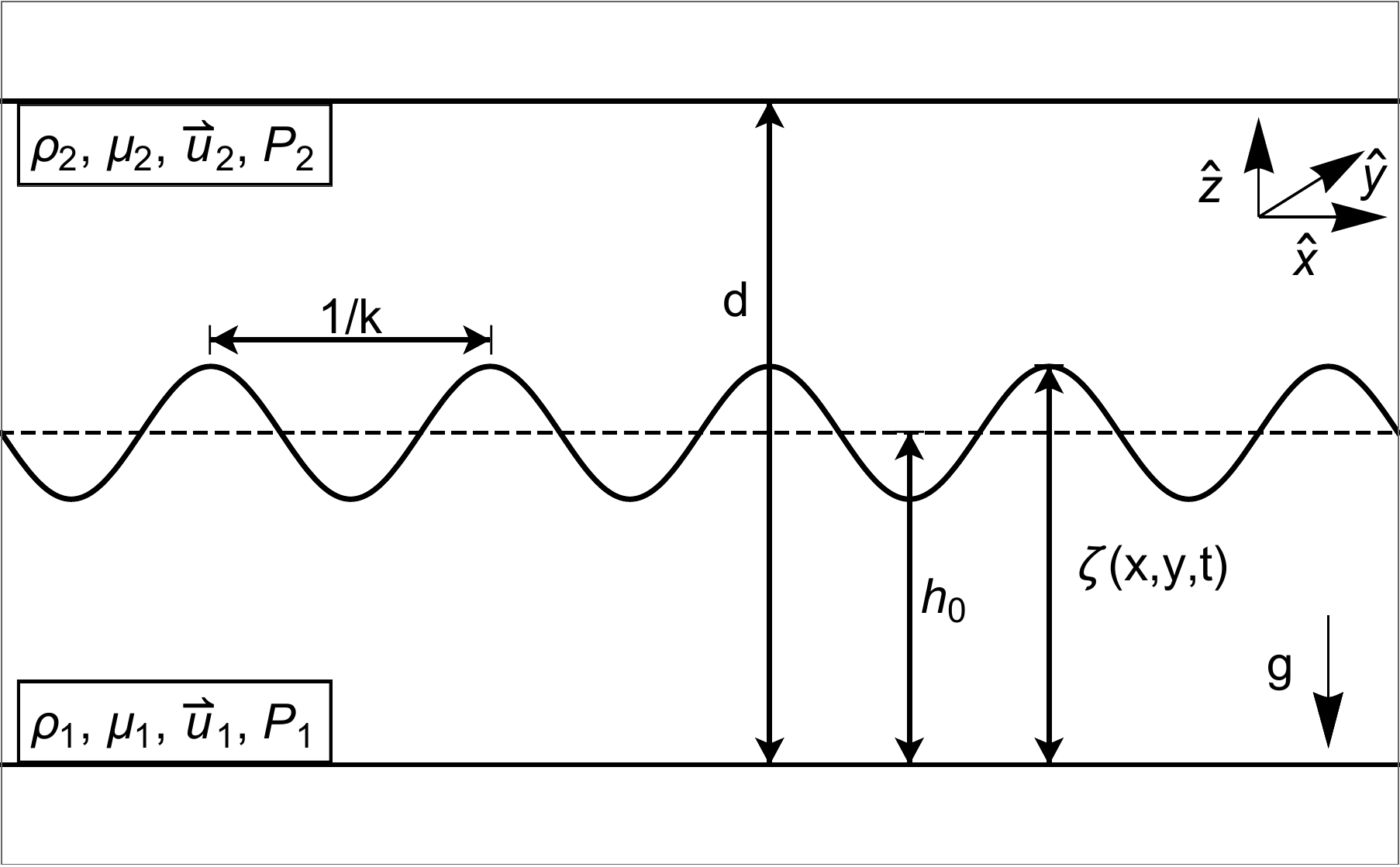}
\caption{The Rayleigh-Taylor instability in the lubrication limit.} 
\label{rtlub}
\end{figure}
It may be asked how stabilization by the Kapitsa effect scales to smaller objects. We therefore consider the situation where the two initially-inert fluids are confined by solid boundaries to a narrow channel of height $d$, each film having a thickness of $h_0$ when the interface is at rest --- see figure \ref{rtlub}. When $d$ is sufficiently large, the results of \S\ref{RTinst} are expected to apply with minor numerical corrections; amending the boundary condition (\ref{nofarflow}) and repeating the analysis leads to the harmonic oscillator-type equation
\begin{equation}
\partial _{tt} \hat{\zeta} = \eta k A g \hat{\zeta},
\end{equation}
where $\eta = 1/\coth(dk/2)$. We are interested in the limit $dk \ll 1$ where $\eta \rightarrow 0$ and the simple harmonic motion seemingly vanishes; this is precisely the lubrication limit. In lubrication-type theories, the unsteady term of the Navier-Stokes equation is typically scaled by $\frac{d^2}{\nu T}$, where $T$ is a representative timescale --- this quantity is then taken to be small \cite{batchelorbook}. For a $\nu$ characteristic of water, using the period $2 \tau$ of the Kapitsa oscillator as a timescale and $\tau \approx 1/60 \ s$ as a representative value, we see that this parameter is small only when $d$ is less than a millimeter. The dynamical equation for the interface in this small-$d$ limit is derived via lubrication theory and then linearized. The details can be found, for example, in \cite{mpbt2005}. In this limit many physical effects are important, however we will only consider gravitational, viscous, and capillary forces. The dynamical equation for a disturbance of the interface is given by
\begin{equation}
\zeta_t = \bar{Q} \left(G(1-\frac{\rho_2}{\rho_1}) \nabla ^2 \zeta - \frac{1}{Ca} \nabla ^4 \zeta \right),
\end{equation}
where $G = (\rho_2 g h_0^2)/\mu_2 u_0$ is the nondimensional gravitation number, $u_0$ is a representative velocity scale, $Ca = \mu_2 u_0 / \gamma (h_0 k)^3$ is the nondimensional capillary number, and $\bar{Q}$ is a factor depending on the viscosities and $d$. Only one derivative of $\zeta$ with respect to time appears; this is due to the absence of the unsteady term $\boldsymbol{u}_t$ which vanishes due to scaling and provides the $\zeta_{tt}$ term in previous sections by differentiating once in time the boundary condition $\boldsymbol{u}\cdot\hat{z}|_{z=0}=\zeta_t$. After the expansion in normal modes, the equation is given by
\begin{equation}
\hat{\zeta}_t = \bar{Q} k^2 \hat{\zeta} \left(G(1-\frac{\rho_2}{\rho_1}) + \frac{k^2}{Ca}  \right). \label{channel1}
\end{equation} 
Applying an identical acceleration as in previous sections allows us to write the gravitational number as a time-dependent quantity, $G(t) = \bar{G}(1+\frac{a}{g \tau^2} \sin (t / \tau))$. In this calculation, unlike our previous calculations, the phase of the oscillator matters and so we consider $G(0)=\bar{G}$ and choose sine over cosine for the form of the acceleration. Unlike Mathieu's equation, (\ref{channel1}) has only one time derivative and admits a solution in terms of well-known functions
\begin{eqnarray}
\hat{\zeta}(t) = e^{\lambda _1 t + \lambda _2 \cos(t/\tau)}, \label{notapend} \\
\lambda_1 = -\frac{\bar{Q}k^2}{Ca} \big[k^2 +  Ca \bar{G}(1-\frac{\rho_2}{\rho_1})  \big], \label{lambda1} \\
\lambda_2 = -\frac{a}{g \tau} \bar{G} k^2 \bar{Q} (\frac{\rho_2}{\rho_1}-1). \label{lambda2}
\end{eqnarray}
It is therefore apparent that when $\lambda_1 > 0$, no value of $\lambda_2$ can cause the interface to be stable for all $t$. The Kapitsa effect, or the presence of strong stabilities of infinite duration guaranteed in the macroscopic system, does not appear in the lubrication limit. In fact, for an injudicious choice of $\lambda_2$, the instability grows much faster than in the absence of the oscillator. However, in actual application the oscillations may be useful in providing short-time stabilities of a denser superposed fluid. Expanding (\ref{notapend}) for small time gives, out to third order, 
\begin{equation}
\hat{\zeta}(t) \approx e^{\lambda_2} \left(1+\lambda_1 t + \frac{t^2}{2}(\lambda_1 ^2 -\lambda_2 / \tau^2) + \frac{t^3}{6}(\lambda_1 ^3 - 3 \lambda_2 \lambda_1 / \tau^2) \right) + ...
\end{equation}
A cursory examination of (\ref{lambda1}--\ref{lambda2}) reveals that for $\rho_2 \gg \rho_1$, $\lambda_2 \approx - \rho_2 / \rho_1 \approx -\lambda_1$, so that the short-time exponential growth is suppressed. For example, when $\lambda_1 \approx - \lambda_2 \approx 1 \ 000$, the disturbance is smaller than a micron for roughly $O(\tau)$ --- this can be a few periods of the oscillator or only the first downstroke in which the film appears to be in free-fall. The proper tuning of $\tau$ in applications of this sort is different than in the macroscopic Kapitsa effect. Decreasing $\tau$ decreases the time the film can be stabilized; increasing $\tau$, on the other hand, increases the necessary amplitude $a$ in order for $\lambda_2$ to remain large and negative. When it is desirable to agitate the interface between two fluids of comparable density, tuning the properties of the oscillator allows one to tune the onset of instability for various wavenumbers. Similar results of enhanced instability for certain material parameters were described by \cite{jl2005} for the related problem of two thin superposed fluid layers flowing down an inclined vibrating plate. Although stabilization can only be achieved for a short time of order $\tau$, which is itself on the order of milliseconds, this still may be meaningful on the timesscales involved in microfluidic devices, where viscous and surface tension effects dominate~\cite{natureMF,tknhcs2014} and systems can reach equilibrium on the order of microseconds. \\

Microfluidic devices display other physical effects which could enhance the short-term stability due to this pseudo-Kapitsa effect. In \cite{mpbt2005} several additional effects have been considered and their inclusion gives rise to a more robust dynamical equation for the interface height. These include an applied voltage $V$, the effects of the Van der Waals forces, and thermocapillary effects at nonzero Marangoni number $Ma$, which is linear in the applied temperature differences across the plates. To get a rough idea of the impact of these effects, we will write the coefficients of $V^2$, $Ma$, and the $d$-dependent Van der Waals effect in a compact notation $e(\varepsilon,d)$, $\Theta(\bar{Q},\lambda,d)$, and $h(H_1,H_2,d)$, where $\varepsilon=\varepsilon_2/\varepsilon_1$ is the ratio of the dielectric permittivities, $H_i$ denotes the Hamaker constant in each fluid region, and $\lambda=\lambda_2/\lambda_1$ is the ratio of the thermal conductivities in the two regions. These dependencies are rather long to write and involve numerical prefactors; the interested reader is encouraged to find the exact dependence on material parameters in the original article. The total dynamics for the interface are then written as
\begin{eqnarray}
\hat{\zeta}_t = \bar{Q} k^2 \hat{\zeta} \big[G(1-\frac{\rho_2}{\rho_1}) + \frac{k^2}{Ca} - e(\varepsilon,d) V^2 + h(H_1, H_2, d) \nonumber \\
+ \Theta(\bar{Q},\lambda,d) Ma \big]. \ \ \ \label{channel2}
\end{eqnarray}
The addition of the new effects is simply quadratic in applied voltage, linear in applied temperature difference, and includes $k$-dependent constants representing gravitation, capillarity, and Van der Waals effects. It is conceivable that the applied voltage or temperature differences could be subjected to oscillations rather than vibrating the whole apparatus, or all three quantities $G$, $V$, and $Ma$ could become time-dependent. Because of the identical linear dependence on both $G$ and $Ma$, inducing a sinusoidal time-dependence in $Ma$ will produce identical results to (\ref{notapend}-\ref{lambda2}) with the inclusion of several new coefficients. Proper tuning of these coefficients may extend the lifetime of the inverted state but do not alter the functional form of the interface height. The quadratic dependence on $V$ is more promising; considering a form $V(t)=V_0\left(1+\cos(t/\tau)\right)$ leads after similar calculations to 
\begin{eqnarray}
\hat{\zeta}(t) = \mathrm{exp}\big[(4 \xi _1 + 6 \xi_2) t + \xi_2 \tau \left(8 \sin\frac{t}{\tau} + \sin \frac{2 t}{\tau} \right)\big], \ \ \ \ \ \ \label{newnotapend} \\
\xi_1 = \bar{Q}k^2 \big[G(1-\rho_2/\rho_1)+k^2/Ca+h(H_1,H_2,d)+Ma\Theta(\bar{Q},\lambda,d) \big], \label{newlambda1} \\
\xi_2 = -\bar{Q}k^2 e(\varepsilon,d)V_0^2. \ \ \ \ \ \   \label{newlambda2}
\end{eqnarray}
Expanding this about $t=0$ gives
\begin{equation}
\hat{\zeta}(t) \approx -(\xi_1 + 4 \xi_2) t + \frac{1}{2} (\xi_1 + 4 \xi_2)^2  t^2 + \big[\frac{2  \xi_2}{\tau^2}-\frac{1}{2}(\xi_1+4\xi_2)^3 \big]t^3 + O(t^4).
\end{equation}
The quantity $(\xi+4 \xi_2)$ appears in the three leading order terms; because the number of parameters inside these two numbers $\xi_i$ is now quite rich, there is a large space of parameter values that will remove this term. The remaining $\sim t^3$ term can then be minimized to effectively remove the three leading order contributions to the interface growth. Choosing reasonable parameters reveals that the inverted state can be sustained until roughly $t \approx \mathcal{O}(10\tau)$. It is reasonable to assume that introducing multiple sinusoidal time-dependences with different timescales could enhance the lifetime of the inverted state even further; however, because this method of inducing stability is different than the Kapitsa effect, we consider further effort to optimize the duration of the inverted state to be outside the scope of the current work.

\section{Instabilities of cylindrical geometry}
\label{cylindersss}

\begin{figure}
\includegraphics[width=\columnwidth]{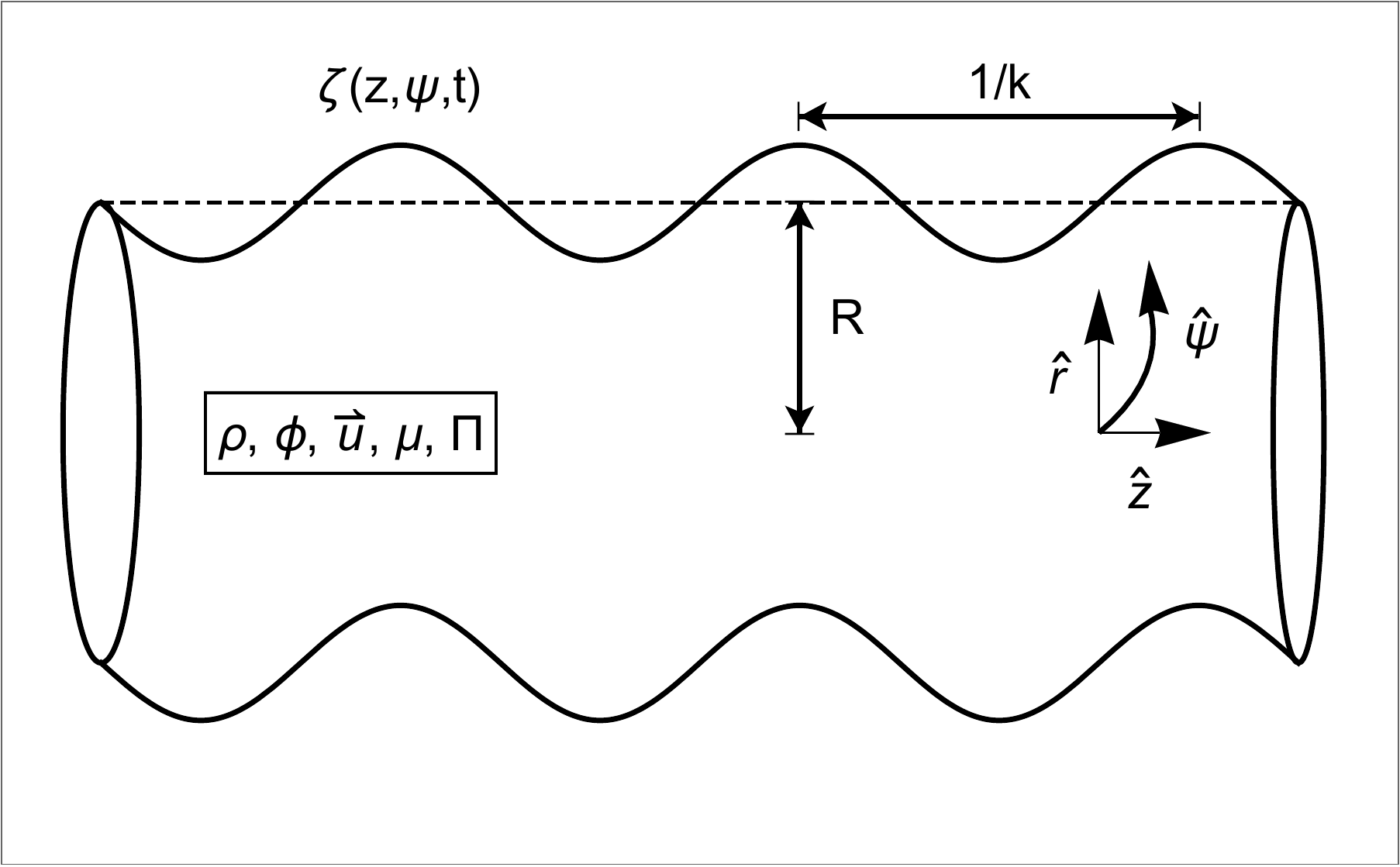}
\caption{Geometry under consideration in the Rayleigh-Plateau and self-gravitational instabilities} 
\label{cylinsetup}
\end{figure}

\subsection{Rayleigh-Plateau instability}
\label{rayleighplateau}
We consider an infinitely long fluid cylinder of radius $R$, density $\rho$ and surface tension strength $\gamma$, which undergoes perturbations of the type $R \rightarrow R + \zeta(t) e^{i(kz + m \psi)}$, where $\psi$ is the azimuthal angle coordinate about the $z$-axis. The system is illustrated in figure \ref{cylinsetup}. For $|\zeta(t)| \ll R$ we can write the governing equations for this system as the gradient of a scaled pressure function $\Pi$, \cite{Chandrasekhar}
\begin{eqnarray}
\boldsymbol{u}_t = - \nabla \Pi, \label{surfcyn1}\\
\nabla \cdot \boldsymbol{u} = 0, \label{surfcyn2}\\
\Pi = \frac{\delta p}{\rho}. 
\end{eqnarray}
Substituting (\ref{surfcyn2}) into (\ref{surfcyn1}) reveals $\Pi$ must satisfy Laplace's equation, which in cylindrical coordinates has the solution
\begin{equation}
\Pi = \zeta(t) \Pi_0 I_m (k r) e^{i(k z + m \psi)}, \label{innerPi}
\end{equation}
where $I_m (b)$ is the pure imaginary Bessel function of the first kind for axial mode number $m$. For a boundary condition we use the Young-Laplace law. In the unperturbed state the inner pressure is given by $p_{in} = \gamma/R$. On the perturbed boundary the condition is $p_{in} + \delta p = \gamma \left( \frac{1}{R_1} + \frac{1}{R_2} \right)$. Substituting the form of the deformation, we recover
\begin{eqnarray}
\boldsymbol{\hat{r}} \cdot \nabla \Pi _{r = R+\zeta} = \zeta \Pi_0 I'_m(k R) e^{i(kz + m \psi)}, \label{surfcyn3}\\
\Pi_0 = -\frac{\gamma}{R^2 \rho} \frac{1-m^2 - k^2 R^2}{I_m(kR)}. \label{surfcyn4}
\end{eqnarray}
Since we are only interested in the dynamics at the boundary, we can examine the dynamical equation (\ref{surfcyn1}) in its neighborhood:
\begin{equation}
\zeta_{tt} = -\boldsymbol{\hat{r}} \cdot (\nabla \Pi)|_{r = R+\zeta}, 
\end{equation}
which, after substitution of (\ref{surfcyn3}--\ref{surfcyn4}), takes the form 
\begin{equation}
\zeta_{tt} = -k \frac{\gamma \zeta}{R^2 \rho} \frac{I'_m(k R)}{I_m (k R)}(1 - m^2 - k^2 R^2). 
\end{equation}
The Rayleigh-Plateau instability does not occur for $m>0$, so we focus on the case $m=0$ \cite{Chandrasekhar}. It is easily checked that if the $m=0$ problem is stable, then the corresponding stability criteria for $m>0$ are also satisfied by basic properties of Bessel functions. Using the relationship $I'_0(x) = I_1 (x)$, we write the dynamical equation for the boundary in pendulum form:
\begin{equation}
\zeta_{tt} = -k \frac{\gamma \zeta}{R^2 \rho} \frac{I_1(k R)}{I_0 (k R)}(1 - k^2 R^2). \label{firstsurfpen1}
\end{equation}
This equation looks like a pendulum in the hanging position for $kR< 1$, which is the parameter region for which the interface is classically stable. For $kR>1$, however, the equation is identical to an inverted pendulum where the ``apparent'' gravitational acceleration is given by $\bar{g} \approx \frac{\gamma}{\rho R^2} \frac{I_1(k R)}{I_0 (k R)}(1-k^2 R^2)$. In the absence of external forces, this configuration will be unstable. 

We now consider a situation in which locally for any value of the azimuthal angle $\psi$ the radius of the cylinder undergoes vibrations of the type $R(t) = R_0 + a \cos (\pi t/\tau)$, performed in such a manner that the total transformations are isochoric. Because we are interested in local dynamics, we ignore the $O(a^2)$ correction to the total volume. We study $\zeta(t)$ in the inertial lab frame and thus add the acceleration of the oscillator according to d'Alembert's principle \cite{landau_lifshitz_mech}. We also consider the leading-order change in the natural acceleration of the system. That is, we expand (\ref{firstsurfpen1}) in $a$ to first order and add the external acceleration, giving
\begin{equation}
\zeta_{tt} = k \zeta \frac{\gamma}{\rho R_0 ^2} \left[ \frac{I_1(k R)}{I_0 (k R)}(k^2 R^2-1) +( \frac{a \rho R_0 ^2}{\gamma}) \cos (\pi t/\tau)  \right], 
\end{equation}

We have again encountered Mathieu's equation. Note that since we are primarily interested in the modes $k R_0 > 1$, we have discarded a term which is exceedingly small in this limit. We can derive the transition to stability for these modes by appealing to Eq. (\ref{arnoldcriterion}): 
\begin{equation}
k a^2 \omega^2 >2 \frac{\gamma}{\rho} (k^2 R_0^2-1)\frac{I_1 (k R_0)}{I_0 (k R_0)} . \label{surfsurfcrit2}
\end{equation}
The right-hand side can be approximated as $k^2 R_0 ^2$  for $k R_0 > 1$, so that the criterion can be written in the convenient form $a^2 \omega^2 > 2 k \gamma/\rho$. The critical wavelength no longer depends on $R_0$. Furthermore, in a fluid such as water, $\gamma/\rho \approx 10^{-5} \ m^3s^{-2}$, so that the right-hand side of (\ref{surfsurfcrit2}) is small. In practical applications the maximum admissible wavenumber $k_{max}$ due to other physical effects, such as the Rayleigh-Taylor mechanism (\S\ref{RTinst}), may be much smaller than the one defined by (\ref{surfsurfcrit2}). Stabilization of the Rayleigh-Plateau problem has been accomplished experimentally by means of acoustic waves which are tuned by use of a camera and computer to have the same wavelength as a developing perturbation of a capillary bridge \cite{mltm1997,mltm2001,bwbd2010}.

\subsection{Invisicid self-gravitating cylinder}
\label{invgalsecc}
We now modify slightly the governing equations of the previous section to reflect an instability instigated by differences in gravitational potential across the boundary instead of pressure differences arising from surface tension:
\begin{eqnarray}
\boldsymbol{u}_t = - \nabla \Pi, \label{invfirsteqngal}\\
\nabla \cdot \boldsymbol{u} = 0, \\
\Pi = - \delta V + \frac{\delta p}{\rho}, \\
\nabla ^2 \delta V = 0,
\end{eqnarray}
where $\delta V$ is the difference in gravitational potential between the inside and the outside of the cylinder; the exact functional forms of both, along with an overview of the problem, can be found in \cite{Chandrasekhar}. The solution of Laplace's equation for $\Pi$ is as before
\begin{equation}
\Pi = \zeta(t) \Pi_0 I_m (k r) e^{i(k z + m \psi)}. \label{invPi}
\end{equation}
A straightforward application of the boundary conditions gives
\begin{equation}
\Pi_0 = -4 \pi G \rho R \frac{K_m (kR) I_m (kR) - 1/2}{I_m (kR)}, \label{invPi0}
\end{equation} 
which has a dependence on $R$ which is much different than that of the Plateau instability. Again we isolate the behavior at the boundary:
\begin{equation}
\zeta_{tt} = -\boldsymbol{\hat{r}} \cdot (\nabla \Pi|_{r = R+\zeta}), \label{invgalloc}
\end{equation}
Keeping terms linear in $\zeta$ and studying the most unstable axial wavenumber $m=0$ reduces (\ref{invgalloc}) to 
\begin{equation}
\zeta_{tt} = \zeta \pi G \rho \big[R k\frac{I_1(k R)(K_0 (kR) I_0 (kR) - 1/2)}{I_0 (kR)}   \big]. \label{invgalfull}
\end{equation}
The nondimensional term in brackets has a root at $k R \approx 1.0668.$ Because $\rho G R$ has units of acceleration, this equation is identical to the equation for a hanging pendulum for $k R > 1.0668$ and an inverted pendulum for $k R < 1.0668$, where the gravitational acceleration is rendered by $\rho G R$ multiplied by the nondimensional geometric factor. 

We now turn our attention to the so-called ``primordial fluctuations'' --- these are fluctuations in the density $\rho$ of a pre-galactic conglomerate of matter to which the formation of some of the universe's earliest structures have been attributed \cite{harrison1970,zeldovich1972}. As recently as two decades ago it was considered that these fluctuations caused instabilities that led to turbulent motion, defining the lengthscales on which galactic structures can be distinguished. However, this was disproved by the experiments of \cite{smoot1992} which determined that the flow in this early epoch was subcritical, with the largest fluctuations falling within the bound $\Delta \rho / \rho \approx 10^{-5}$. A more detailed account of this field can be found, for example, in \cite{gibsonPhD}. This led to the development of several theories seeking to explain why the self-gravitational instabilities did not occur. Here we consider a toy problem in which the instability is stabilized by vibrations.

We do not speculate on the mechanism by which the mass of an entire proto-galaxy can be made to vibrate, but simply observe that in the astophysical literature the primordial density fluctuations were accompanied by acoustic phonons due to \textit{metric} fluctuations \cite{zeldovich1972}. These metric fluctuations were distinguished by a scale-free conformal parameter $b \approx 10^{-4}$. We naively study the equation (\ref{invgalfull}) under the transformation $R(t) = R_0 e^b \cos (\pi t/\tau) \approx R_0 + b R_0 \cos (\pi t/\tau)$. We now rewrite (\ref{invgalfull}) as Mathieu's equation,
\begin{eqnarray}
\zeta_{tt} = \pi \zeta \rho G R_0 k \big[\frac{I_1(k R_0)(K_0 (kR_0) I_0 (kR_0) - 1/2)}{I_0 (kR_0)} \nonumber \\
+ ( \frac{b }{\rho G}) \cos (\pi t/\tau)    \big]. 
\end{eqnarray}
 
The stability criterion (\ref{arnoldcriterion}) becomes

\begin{equation}
b^2 \omega^2 > 2 \rho G \frac{I_1(k R_0) \left(K_0(k R_0) I_0 (k R_0) - 1/2\right)}{kR_0 I_0(k R_0)}, \label{invgalfinal}
\end{equation}
the right-hand side of which can be approximated by $-0.19-\log (k R_0)/2$ in the regime of interest, $k R_0 < 1$. This has a pole at $k R_0=0$, so that not all modes will be stable, but the stability boundary will be pushed well past $k R_0=1$ depending on the value of $\rho$. Between the times when the universe was approximately $10^2$ seconds old and $50 \ 000$ years old, $\rho G/b^2$ varied between approximately $1$ and $10^{-19}$ \cite{smoot1992}. In our toy model the requisite period of the oscillator to induce stabilization therefore varies between the timescales of seconds and millenia, depending on the value of $\rho$.

We conclude this section with some notes on the possible effects of both viscosity and a magnetic field on this calculation. Experimental studies consistently reveal that the universe is magnetized on several lengthscales. Several mechanisms have recently been proposed to explain how the presence of a magnetic field on all probed lengthscales might have evolved from a primordial magnetic field; for a recent comprehensive review, see \cite{subramanian}. Glossing any relevant text \cite{Chandrasekhar} reveals that the role of an axial magnetic field in an inviscid cylindrical proto-galaxy has a stabilizing effect which is entirely analogous to that of surface tension in the Rayleigh-Taylor instability. In this context Eq. (\ref{surteneqn111}) suggests that stability is enhanced by a static field. This also suggests a route to global stability in the case where there are no vibrations, but rather small-amplitude high-frequency oscillations about an average value for this magnetic field --- this can stabilize against all disturbance wavenumbers even for a relatively weak field average. Including the effects of viscosity also transforms this harmonic oscillator in a damped one in analogue with the Rayleigh-Taylor instability; depending on the epoch under consideration for the early universe, the viscosity may actually have been quite large due to the interaction of photons and the primordial plasma, and may need to be included.

\section{Conclusion}
\label{conclusions}
In this work we reviewed existing literature connecting an inverted pendulum to the Rayleigh-Taylor instability. Both systems can be stabilized by low-amplitude high-frequency external forcing in the vertical direction --- mathematically speaking, this equates to a transformation from the equation of a simple harmonic oscillator to Mathieu's equation. We expand this initial correspondence by deriving similar equations for discrete planar and cylindrical interfacial instabilities of interest and pointing to further expansions known in the literature. In each case we use the pendulum analogies to invoke results from the vibrations and dynamical systems literature. In the two cylindrical interfaces studied --- those governed by the Plateau and self-gravitational instability, respectively --- the apparent gravitational acceleration of the analogous pendulum is extremely small, suggesting that the instability is easily suppressed by vibrations. We speculate on some possible mechanisms by which the effect can be generated in these geometries. \\

 \bibliography{newrefs}

\end{document}